\documentclass[sigconf]{acmart}
\pdfoutput=1
\AtBeginDocument{%
  \providecommand\BibTeX{{%
    \normalfont B\kern-0.5em{\scshape i\kern-0.25em b}\kern-0.8em\TeX}}}

\usepackage{xspace}
\usepackage{balance}
\usepackage{graphicx}
\usepackage{booktabs}

\newcommand{\ie}{\emph{i.e.,}\xspace}
\newcommand{\eg}{\emph{e.g.,}\xspace}

\newcommand{\popmethod}{\textit{MostPop}\xspace}
\newcommand{\paratitle}[1]{\vspace{1.5ex}\noindent \textbf{#1}}

\copyrightyear{2020}
\acmYear{2020}
\setcopyright{acmcopyright}
\acmConference[SIGIR '20]{Proceedings of the 43rd International ACM SIGIR Conference on Research and Development in Information Retrieval}{July 25--30, 2020}{Virtual Event, China}
\acmPrice{15.00}
\acmDOI{10.1145/3397271.3401233}
\acmISBN{978-1-4503-8016-4/20/07}

\begin{document}
\fancyhead{}

\title{A Re-visit of the Popularity Baseline in Recommender Systems}

\author{Yitong Ji, Aixin Sun,  Jie Zhang}
\affiliation{%
  \institution{Nanyang Technological University, Singapore}
}
\email{{yitong.ji, axsun, zhangj}@ntu.edu.sg}

\author{Chenliang Li}
\affiliation{%
  \institution{Wuhan University, China}
}
\email{cllee@whu.edu.cn}

\begin{abstract}
Popularity is often included in experimental evaluation to provide a \textit{reference performance} for a recommendation task. To understand how popularity baseline is defined and evaluated, we sample 12 papers from top-tier conferences including KDD, WWW, SIGIR, and RecSys, and 6 open source toolkits. We note that the widely adopted  \textit{MostPop} baseline simply ranks items based on the number of interactions in the training data. We argue that the current evaluation  of popularity (i) does not reflect the popular items at the time when a user interacts with the system, and (ii) may recommend items released after a user's last interaction with the system. On the widely used MovieLens dataset, we show that the performance of popularity could be significantly improved by 70\% or more, if we consider the popular items \textit{at the time point} when a user interacts with the system. We further show that, on MovieLens dataset, the users having lower tendencies on movies tend to follow the crowd and rate more popular movies. Movie lovers who rate a large number of movies, rate movies based on their own preferences and interests. Through this study, we call for a re-visit of the popularity baseline in recommender system to better reflect its effectiveness.
\end{abstract}


\maketitle

 \section{Introduction}
\label{sec:Introduction}

Recommender systems aim to predict user preferences of items, hence to present a user with items of her interest. Various different models have been proposed, from non-personalized method like "Most Popular" (\popmethod) to personalized solutions. \popmethod, arguably the simplest recommendation method, recommends items with high popularity.  As an easy to implement and non-personalized recommendation method, \popmethod is widely used as a baseline to provide a \textit{reference performance} for a recommender system.

The evaluation of model effectiveness can be conducted online or off-line.  As many researchers do not have access to real platforms for online evaluation, off-line evaluation is widely adopted. Depending on how a dataset is partitioned into training and test sets, the evaluation conducted on test set may not be realistic in online setting. In this study, we sample 12 papers from top-tier conferences and 6 toolkits to have a better understanding of the \popmethod. We observe that many papers and even toolkits fail to provide proper definitions of the \popmethod method. The evaluation conducted, therefore may not truly reflect the effectiveness of the popularity baseline. A key issue here is the ignorance of the time dimension in popularity definition and evaluation.

In our experiments conducted on the widely used MovieLens dataset, we show that, by simply considering the time dimension in popularity definition and evaluation, the performance of the popularity baseline could be improved by 70\% or more, measured by Hit Rate (HR) and Normalized Discounted Cumulative Gain (NDCG). Note that, the term ``time dimension'' in our discussion is different from that in time-aware recommender systems. In our discussion, the time dimension means that popularity of an item is defined with respect to time (see Figure~\ref{fig:popExp}). On the same dataset, we also show that, popularity is more effective for users who do not have many interactions with the system.  Next, we discuss the problems with the mainstream definition and evaluation of the popularity baseline, and then evaluate two new popularity methods, named \textit{RecentPop} and \textit{DecayPop}. Lastly, we study the effectiveness of \textit{DecayPop} on different groups of users.

\section{Popularity Definition}
\label{sec:definition}

To understand how ``popularity'' is defined and evaluated, we sample papers from top tier venues including SIGIR, KDD, WWW, and RecSys.\footnote{\scriptsize It is not our interest to provide a comprehensive survey of definition and evaluation of popularity method. We believe the sampled papers give a good representation of the understanding of \popmethod.} All the sampled papers use MovieLens dataset in their evaluation and all treat the dataset as an implicit feedback dataset. That is, a rating by a user to a movie is an indication of a user-item interaction. A recommendation method predicts whether a user interacts with a movie and ignores the actual value of the rating.  We sample papers by usage of the MovieLens dataset because we use the same dataset in our analysis. More importantly, MovieLens dataset provides timestamps of user-item interactions (\ie the time point when a user rates a movie).

\paratitle{What is \popmethod?} As the simplest baseline, many studies only give brief definition of \popmethod as ``a non-personalized method'' and  state that ``popular items would be recommended''~\cite{Wu2019,Chen2019,Otunba2017, Palumbo2017, Sun2018, Pasricha2018, Canamares2019}. There are also works which highlight that popularity can be measured by using the number of explicit/implicit feedbacks on items~\cite{Dacrema2019, Du2019, Balog2019,Yuan2019, He2017}.  We also note that, popularity is defined by ``number of interactions in training data'' in a number of open-source recommendation tools, including LibRec~\cite{Guo2015}, Sequence-Based-Recommender~\cite{Rec_with_RNN}, RecommenderLab~\cite{Hahsler2011}, CaseRecommender~\cite{caseRecommender}, Microsoft Recommender~\cite{Graham2019} and TagRec~\cite{Kowald2017}. To summarize, many studies define and evaluate ``popularity'' as the \textit{overall popularity} of items in training set. In other words, the items to be recommended to users are those with the highest number of interactions in the training data. Depending on how the off-line data is partitioned into training and test sets, we argue that in many  existing evaluations, \popmethod does not truly reflect our common understanding of popularity.

\begin{figure}
	\centering
	\includegraphics[trim=3cm 6.8cm 9cm 4cm, clip, width=0.8\columnwidth]{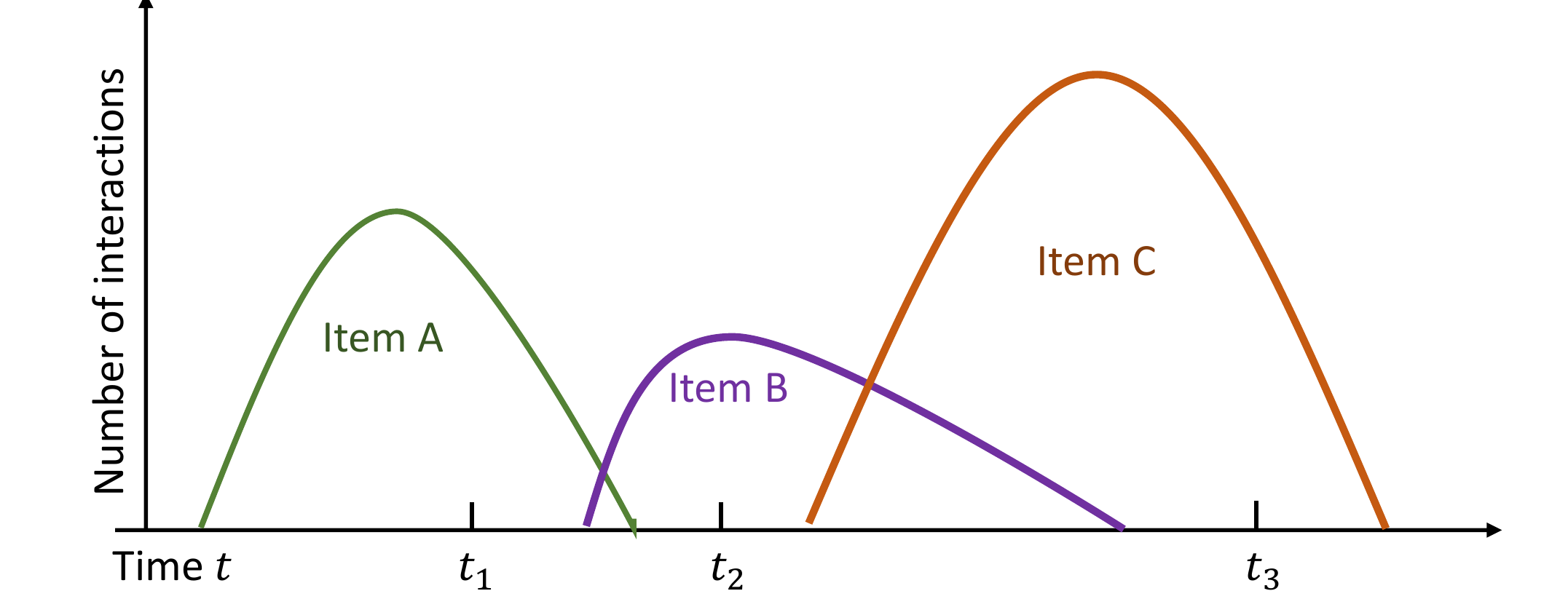}
	\caption{Popularity of items $A$, $B$, and $C$ over time.}
	\label{fig:popExp}

\end{figure}

Typically a long time period is needed to accumulate a reasonably large number of interactions between users and items for off-line evaluation. The popularity of items may change from time to time. We illustrate this point in Figure~\ref{fig:popExp} with three example items $A$, $B$, and $C$ and their number of interactions over time. If a user interacts with the system at $t_1$, $t_2$, and $t_3$, then the \textit{current} ``most popular'' items will be  $A$, $B$, and $C$ respectively, assuming there are only 3 items in the system. However, if we take the overall popularity, then the ranking becomes $C$, $A$, and $B$ in the entire dataset.

\paratitle{How to evaluate \popmethod?}  The example in Figure~\ref{fig:popExp} also brings in another interesting problem on the validity of the off-line evaluation. In off-line evaluation, there are multiple ways to partition the data into training and test sets, \eg randomly sampling 80\% of $\langle$user, item, interaction$\rangle$ tuples as training and the remaining as test~\cite{Chen2019,Otunba2017}; taking 80\% of interactions per user as training and the remaining as test~\cite{Wu2019,Palumbo2017,Sun2018}; or leave-one-out by taking the last interaction of each user as test and previous interactions as training~\cite{Yuan2019,He2017,Pasricha2018}. We use leave-one-out to illustrate the point.

As off-line data is accumulated over a time period, the last interaction of a user may appear at any time point along the time period. For example, in Figure~\ref{fig:popExp}, we may consider $t_1$ to be the last time point user $u_1$ interacts with the system, and $t_2$ and $t_3$ be the time points of last interactions from $u_2$ and $u_3$ respectively. Then regardless how popular items $B$ and $C$ are, $u_1$ has never had the chance to access these items, because these items are not released yet at time $t_1$. However, with leave-one-out partition, as existing \popmethod simply ranks items based on accumulated user interactions in training set, then items released after a user's last interaction could be recommended to this user, which is not feasible in reality.

\begin{figure}
	\centering
	\includegraphics[ width=0.85\columnwidth, scale= 0.8]{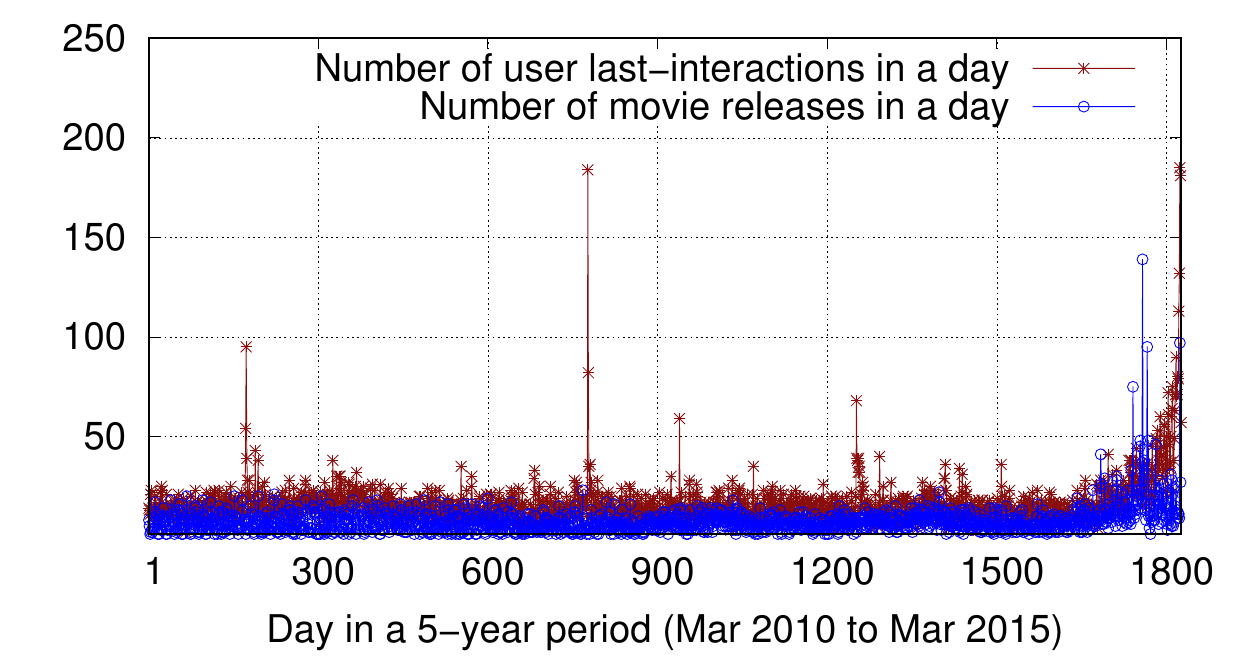}
	\caption{Number of  movie releases and number of user last-interactions on each day in MovieLens (best viewed in color)}
	\label{fig:userMovie}
\end{figure}

To further illustrate this point, we conduct a simple experiment on MovieLens dataset. We consider that a movie is released (or available to users) from the time when its first rating is received from any user. Then we follow leave-one-out to partition the data into training and test sets; the last-interaction from each user will be in the test set. We plot the number of user last-interactions and number of movie releases on each day, in Figure~\ref{fig:userMovie}. It shows that both user last interactions and movie releases well spread along the time line. If we consider a user leaves the system after her last interaction, then many users have no access to many movies released after. Because evaluation reflects how good a recommender method recovers user-item interactions in test set, then recommending movies released after a user leaves the system will never lead to a hit. The \popmethod method that simply ranks all items in training set without considering the time dimension, will lead to poor performance of the ``popularity'' baseline.

\section{Most-Pop, Recent-Pop, and Decay-Pop}
\label{sec:popularity}

The \popmethod definition does not reflect our common understanding of popularity (an item is usually popular for a time period as shown in Figure~\ref{fig:popExp}) and may not be evaluated correctly. In this section, through experiments, we demonstrate that by simply taking time into account, the accuracy of recommendation by popularity could be improved significantly.

We conduct experiments on MovieLens20m dataset which covers 20 years of interaction data (09 Jan 1995 to 31 Mar 2015). By leave-one-out, we hold out the last interaction of each user as test and the remaining interactions as training, the same as  mainstream leave-one-out setting. To simulate the situation that a system requires a long time to accumulate user interactions, we evaluate the models by using the test instances that happened in the last five years (31 Mar 2010 to 31 Mar 2015). As the result, we have 29,431 test instances in our evaluation. Hit Rate (HR) and Normalized Discounted Cumulative Gain (NDCG) are used as the evaluation metrics. We compare the following three "popularity" models.

\paratitle{MostPop.} This is the mainstream popularity baseline widely seen in research papers and toolkits. MostPop ranks items by the number of interactions in the entire training set.

\paratitle{RecentPop.} RecentPop is designed to recommend the popular movies to a user at the time point when she interacts with the system, \ie timestamp of her last interaction in testing, denoted by $t_0$. We derive the most popular movies at time $t_0$. As we discussed earlier, a movie could be popular for a short time period and then become less popular. For this reason, we rank the movies by their number of ratings received within a short time period $[t_0-\Delta t, t_0]$. $\Delta t$ is set to 1 month in our evaluation.

\paratitle{DecayPop.} DecayPop is a simple extension of RecentPop, to derive popular movies in a longer time period before $t_0$. We consider the past 6 months before $t_0$ and take the weighted sum of number of ratings in the 6 months. Higher weight is assigned to more recent interactions of movies. Specifically, the number of ratings in each month is weighted by an exponential decay $e^{-t_m}$. Here, $t_m\in [1, 6]$ is the number of months with respect to $t_0$ and $t_m=1$ for the most recent month.

Both RecentPop and DecayPop remain non-personalized methods, the same as MostPop. The only difference is that RecentPop and DecayPop consider the time point when a user interacts with the system, and derive the most popular items at that time point.

\begin{table}
\centering
		\caption{Results of popularity methods; best results in bold}
		\label{tab:mainPop}
		\begin{tabular}{l|cccc}		
				\toprule
				\textbf{Popularity} & \textbf{HR@5} & \textbf{HR@10} &  \textbf{NDCG@5} &  \textbf{NDCG@10}\\
				\midrule
				MostPop & 0.0304 & 0.0462 & 0.0198 & 0.0248\\
				RecentPop & 0.0530 & \textbf{0.0845}  & 0.0338 & 0.0440\\
				DecayPop& \textbf{0.0532} & 0.0843& \textbf{0.0341} & \textbf{0.0441}\\	
				\bottomrule		
		\end{tabular}
\end{table}

Table~\ref{tab:mainPop} reports  recommendation accuracy of the three popularity methods. We make two observations. First, RecentPop achieves significant improvement over MostPop on all evaluation metrics. The improvement is in the range of  70\% to 83\%.  DecayPop further improves the recommendation accuracy by a small margin and achieves the best results on three out of four measures.

By taking into account time dimension in evaluation, RecentPop and DecayPop achieve significant improvement over MostPop. Both new popularity definitions are extremely simple to implement (and also customizable by adjusting $\Delta t$ and weighting function) . We argue that both RecentPop and DecayPop better reflect a reference performance than MostPop.

\section{Movie Popularity vs User Activity }
\label{sec:userBehaviour}

Some users are more active than other users by the number of interactions in a system. For an item to be popular, it must receive a large number of votes from many users. An interesting question here is who these users are: \textit{are they the users who interact a lot with the system, or common users who are not very active?}

\begin{figure}
	\centering
	\includegraphics[scale=0.5]{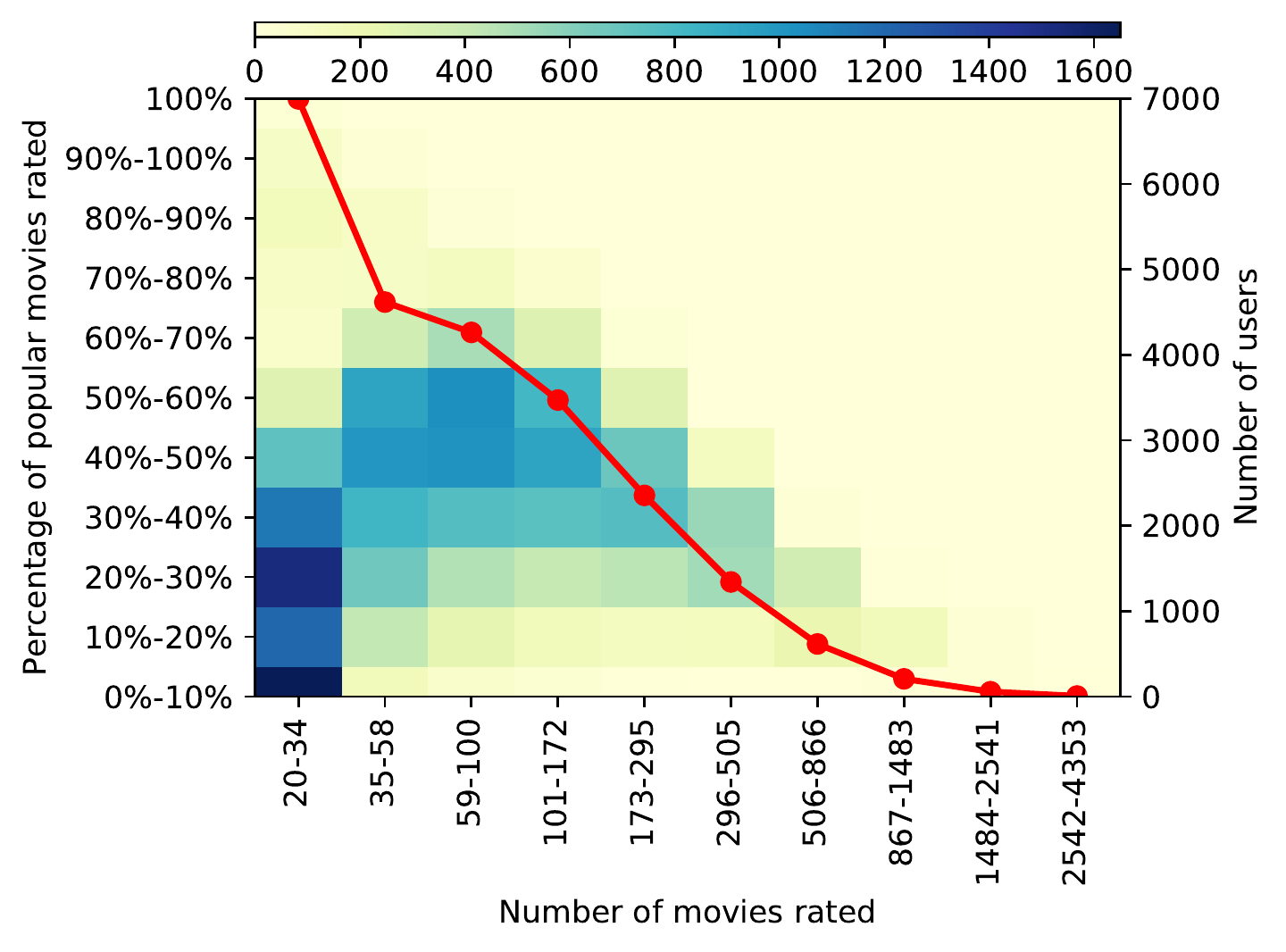}
	\caption{Users are partitioned into groups based on the number of movies rated ($x$-axis). The line in red plots the number of users in each user group, w.r.t $y$-axis on the right hand side. Heatmap shows the number of users with the specified percentage ($y$-axis on the left hand side) of popular movies among all her rated movies, in a user group.}
	\label{fig:perPopMovies}
\vspace{-2ex}
\end{figure}

To study how users interact with popular movies, we conduct an analysis on MovieLens20m.  In this analysis, we only consider the users who have their first movie ratings happened in a 5-year period (31 Mar 2010 to 31 Mar 2015), resulting in 23,934 users. By log-scaled number of movies rated in this 5-year period, we partition these users into 10 groups. In Figure~\ref{fig:perPopMovies}, the line with dots, with respect to the $y$-axis on the right hand side, shows the number of users who fall in each group, with the number of movies rated along the $x$-axis.\footnote{\scriptsize MovieLens data filters users who have fewer than 20 ratings.} As expected, a large number of users rate a small number of movies and a small number of users rate a few thousands of movies.

\paratitle{Tendency to popular movies.} If a user rates $x$ number of movies in total, we are interested to know, among $x$, what the percentage of ``popular movies'' is. For this reason, we define ``popular movies'' in this analysis to be the top-200 ranked movies by DecayPop at any time point of interest. The heatmap in Figure~\ref{fig:perPopMovies} reports the number of users who fall in each cell, where $x$-axis is the user group and $y$-axis (on the left hand side) is the percentage of popular movies among all movies rated by a user. The list of popular movies is derived at the time point of user rating. The first user group (having 20-34 movies rated in total) typically have 0-10\% and 20\%-30\% of popular movies among their rated movies. The second user group (having 35-58 movies rated in total) typically have 30\%-60\%  popular movies among all their ratings. To put these percentages in context, we compute the chance of selecting a popular movie from all movies at random. On the first day of the 5-year period, the number of movies already available in the dataset is 13,415, and the number of movie grows over time (see Figure~\ref{fig:userMovie} for new movie releases). The chance of picking up a popular movie at random at any time point is smaller than $200/13,415\times 100\% = 1.49\%$. Figure~\ref{fig:perPopMovies} shows that most users who rate fewer than 295 movies have high tendencies to rate popular movies. We note that if a user rates more than 200 movies, then the percentage of popular movies among all rated movies will reduce along the increase of rated movies (\ie denominator increases). Nevertheless, the number of users who rate a large number of movies is small, and very few users fall in the last three user groups.

\begin{figure}
	\centering
	\includegraphics[width=0.9\columnwidth]{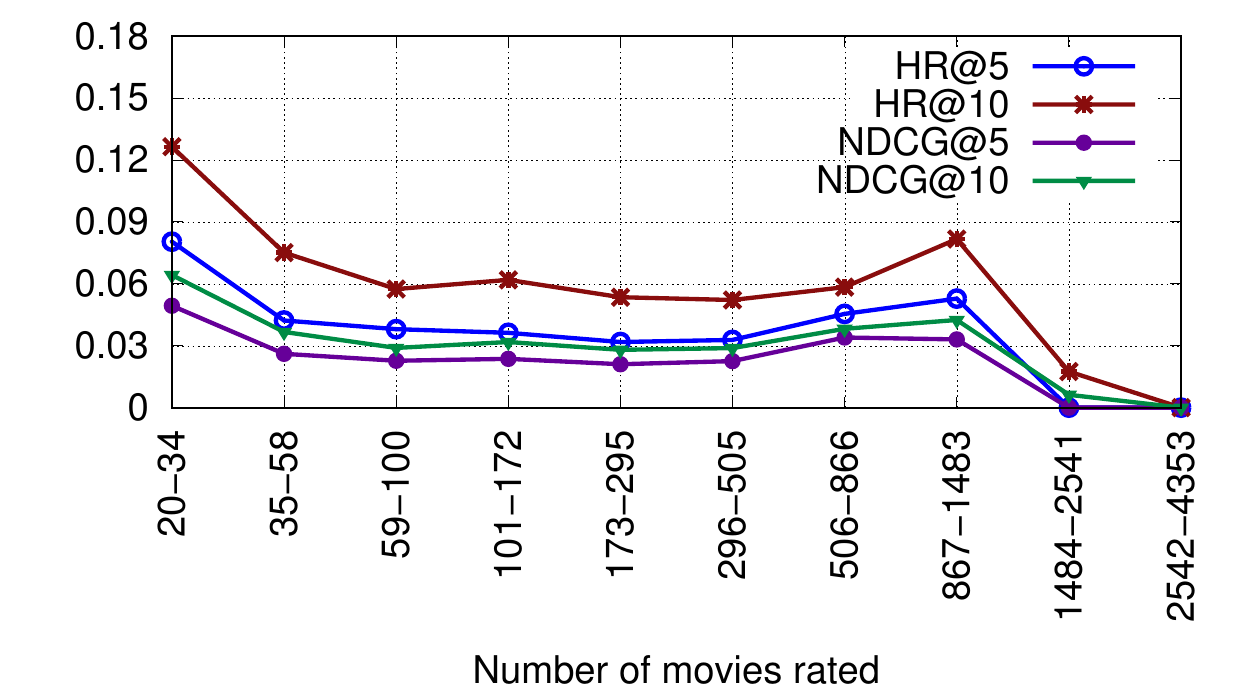}
	\caption{Performance of DecayPop by user groups.  }
	\label{fig:performance}
\vspace{-2ex}
\end{figure}

\paratitle{Recommendation accuracy by user group.} We further test recommendation accuracy of DecayPop on the aforementioned 10 user groups. Figure~\ref{fig:performance} reports the performance. Consistent trend is observed on all evaluation metrics. In Figure~\ref{fig:perPopMovies}, we note that percentage of popular movies for user group in the range of 20-34 is less than 40\%, which is  lower than other user groups (\eg user group 35 - 58). However, the best performance is obtained for the 20-34 group in Figure~\ref{fig:performance}. The reason is that Hit Rate and NDCG are computed for top-5 and top-10 ranked movies, and popular movies in Figure~\ref{fig:perPopMovies} are computed based on top-200 movies. The high values for the user group 20-34 suggest that, this group of users are more likely to rate the most popular movies. Note that this group is also the largest group of users (see Figure~\ref{fig:perPopMovies}).  We observe  that, the lines in Figure~\ref{fig:performance} demonstrate small spikes for the last three or four user groups. Because the number of users in these user groups are very small, the results may not show a smooth trend.

Our results suggest that when a user does not rate movies often, then she is more likely to follow the crowd and rate popular movies; if a user rates many movies, then she tends to follow her own interests to choose movies. In this sense, popularity could be a strong baseline for users who do not have many interactions with a system, if defined and evaluated properly.

\section{Related Work}
\label{sec:related}
Studies on the popularity method itself are very limited. The most relevant work to ours is Time-Aware Local Popularity, TimePop in short~\cite{Anelli2018}. TimePop extracts neighbors who have similar historical interactions as target users. It then considers the time point at which the neighbors rated the items. Decay function is then applied to ratings with the consideration of time, so that items rated more recently receive higher weights. The weighted sum of interactions is used to measure items' popularity. TimePop does consider time in its computation. However, it only ranks items that are rated by target users' neighbors. In this sense, TimePop is a personalized method as different users' neighbors are different. In our study, we challenge the definition and evaluation of the mainstream popularity baseline, which is non-personalized. Software popularity algorithm based on evaluation model~\cite{Wang2018} is a non-personalized method to recommend popular softwares to users. In the computation of popularity, time duration since an item's release date is applied as a shrinkage factor to popularity. Hence recently released items have a higher chance of being recommended. Again, in our study, we focus on the definition and evaluation of the mainstream popularity baseline.

We note that our study is also different from the line of work on the bias of popularity on recommender systems~\cite{Canamares2019}. This line of studies shows that some of the current recommender systems tend to recommend popular items.

\section{Conclusion}
\label{sec:Conclusion}
In this paper, we focus on the simplest recommendation baseline popularity. We note that the mainstream popularity definition does not well reflect  our common understanding that popularity of items changes over time. We also argue that, the evaluation of popularity in most existing settings may not truly reflect its effectiveness. By simply considering the time dimension, we show that RecentPop and DecayPop could improve the recommendation accuracy by over 70\% compared to \popmethod. Our further analysis shows that recommendation by popularity is effective for users who do not have many interactions with a system. Through this study, we call for a re-visit of the popularity baseline. As a simple non-personalized baseline, popularity shall be evaluated in a more proper manner, to provide meaningful reference performance in recommendation tasks. As a part of our future work, we will study the evaluation of other recommendation models in off-line settings.

\section{Acknowledgments}
This work was conducted within the Delta-NTU Corporate Lab for Cyber-Physical Systems with funding support from Delta Electronics Inc. and the National Research Foundation (NRF) Singapore under the Corp Lab@University Scheme.

\bibliographystyle{ACM-Reference-Format}
\balance
\bibliography{references}

\end{document}